\documentclass[prx,aps,twocolumn, reprint]{revtex4-2}
\usepackage{graphicx}
\usepackage{aurical}
\usepackage{amsmath, amssymb}
\usepackage[T1]{fontenc}
\usepackage[version=3]{mhchem}
\usepackage{braket}
\graphicspath{{./figures/}}
\DeclareGraphicsExtensions{.png,.pdf}

\usepackage{silence}
\newcommand{\una}{Universit\'{e} de Nouakchott, Facult\'{e} des Sciences et Techniques, D\'{e}partement de Physique, Avenue du Roi Fai\c{c}al, 2373, Nouakchott, Mauritania}

\usepackage[pdftex, colorlinks=true, linkcolor=blue, citecolor=blue, urlcolor=blue]{hyperref}

\begin{document}
\title{High-Harmonic Spin and Charge Pumping in Altermagnets}

\author{O. Ly}
\email{ousmanebouneoumar@gmail.com}
\affiliation{\una}

\begin{abstract}
We report the emergence of highly nonlinear spin and charge pumping in an altermagnetic system driven by magnetic dynamics. The non-relativistic spin–momentum coupling inherent to altermagnets (AMs) generates a giant momentum dependent spin splitting, leading to strong spin-flip scattering in the presence of a precessing magnetic order driving the altermagnetic system out of equilibrium. Our simulations reveal the emission of hundreds of harmonics under realistic conditions, with amplitudes far exceeding those obtained in light-driven schemes. Notably, in contrast to ferromagnetic and conventional antiferromagnetic systems, where nonlinear emission typically requires additional spin–orbit coupling, AMs intrinsically support high-harmonic spin and charge pumping. These results identify altermagnetic systems as a promising platform for efficient THz emitters and highly nonlinear spintronic devices.
\end{abstract}

\maketitle

\section{Introduction}
Light-matter interaction has long been a central topic in condensed matter physics
\cite{HaugKoch2009,Flick2017,Kockum2019,Hubener2017}. A particularly ubiquitous phenomenon that arises from this interaction is the generation of harmonics, which originates from the intrinsically nonlinear response of matter to strong electromagnetic fields \cite{Corkum1993,Vampa2015}. Since the early 1990s, high harmonic generation (HHG) has been extensively investigated in molecular and atomic systems, where nonlinear emission is commonly observed in gaseous media \cite{Krause1992,Huillier1993,Lewenstein1994}. The ultrahigh-harmonic regime was subsequently demonstrated using intense laser fields interacting with high-pressure gasses \cite{Popmintchev2012}. These developments ultimately led to the emergence of atto-second science, enabling the investigation of electron dynamics on ultrashort time scales \cite{Vampa2017,Li2020,Uzan2022,Inzani2025}.

{Furthermore, a plethora of condensed matter systems have been investigated in this perspective, including strongly correlated electrons \cite{Murakami2018,Imai2020}, non-collinear magnetic textures \cite{Ono2024} and other topological systems \cite{Baykusheva2021,Garcia-Cabrera2024,Venkat2026}. This establishes HHG as a key concept for condensed matter photoemission spectroscopy \cite{Zhong2022} with emerging new frontiers in ultrafast quantum light science \cite{Ciappina2025}.}

Only recently have alternative routes to ultrafast {carrier} dynamics been proposed, in which magnetic order, rather than light, serves as a driving mechanism. In particular, magnetization dynamics in systems with strong spin-orbit coupling (SOC) has been shown to induce highly nonlinear spin and charge transport responses \cite{Ly2022}. This approach has been explored in Rashba spin-orbit-coupled systems as well as in non-collinear antiferromagnets hosting topological real-space textures \cite{Ly2023}. In these systems, the coupling between the vectorial dynamics of the magnetic order and the electronic spin degrees of freedom enables magnetically driven HHG. Importantly, this mechanism is expected to operate in both relativistic and non-relativistic spin-orbit-coupled materials.

A particularly promising platform in this context is the recently proposed altermagnetic phase, which is characterized by a giant, momentum-dependent spin splitting despite the absence of net magnetization. This phase has been theoretically predicted \cite{Satoru2019, Kyo-Hoon2019, Satoru2020, Yuan2020, Smejkal2020, Smejkal2022a} and experimentally realized in several materials \cite{Fedchenko2024,Reimers2024,Jianyang2024, Yang2025,Yang2025b}. Altermagnets (AMs) have also been shown to exhibit rich transport properties \cite{Weber2024,Weber2025,Fu2026,Zhou2026} when interacting with light, making them attractive candidates for applications in ultrafast spintronics and orbitronics.

So far, recent theoretical work \cite{Sun2023,Hodt2024} has begun to explore spin pumping in AMs beyond the conventional ferromagnetic and antiferromagnetic paradigms. These studies focus on regimes of weak or effectively perturbative magnetic dynamics such as small angle precession or linear response in which strongly nonlinear behavior and HHG are not pronounced.

In the present work, we perform exact (non-perturbative) non-equilibrium quantum transport calculations \cite{Kloss2021} to reveal high-harmonic spin and charge dynamics in an altermagnetic system, employing a realistic model that captures the essential features of the momentum-dependent spin splitting reported in first principles band-structure calculations \cite{Roig2024,Kyo-Hoon2019}. 
Building on our recent proposal that nonlinear transport can be anticipated from the dynamics of the instantaneous energy levels \cite{Ly2025}, we analyze the energy level evolution under different excitation geometries and reveal pronounced high-harmonic emission arising from the interplay between spin-polarized altermagnetic bands and a precessing magnetic order. Importantly, in contrast to light-driven HHG in AMs \cite{Werner2024}, where higher harmonics typically appear with relatively small amplitudes, our simulations indicate that magnetically driven dynamics can generate harmonics extending well beyond the $100$th order, with amplitudes comparable to the fundamental response. These findings highlight magnetic dynamics as a powerful and efficient route to strong nonlinear emission in non-relativistically spin-split materials.

\begin{figure*}
    \includegraphics[width=0.8\textwidth]{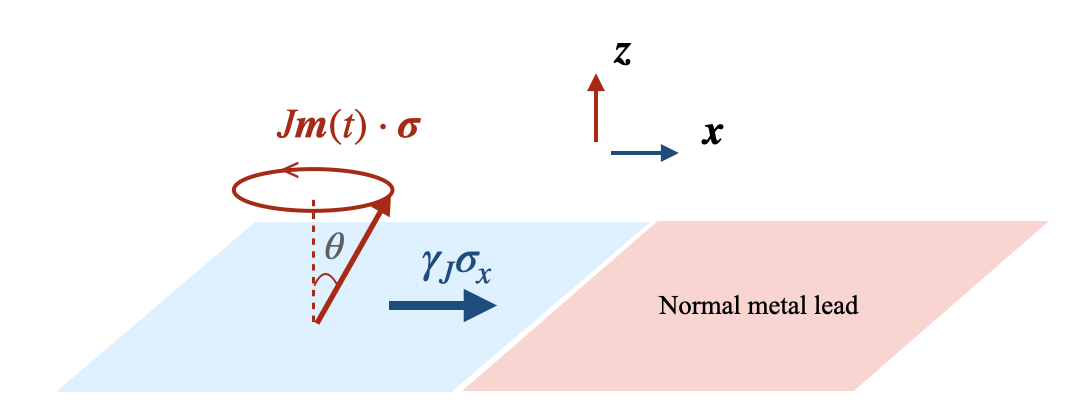}
    \caption{Schematic of the magnetic configuration employed in the numerical simulations. A time-dependent magnetic order precesses around the 
$z$ axis (red cone) in the presence of an altermagnetic order parameter oriented along the 
$x$ axis (blue arrow). The magnetic scattering region (blue rectangle) is connected to a normal-metal lead (red rectangle), where the resulting time-dependent carrier currents can flow.
}
    \label{fig:fig1}
\end{figure*}

\section{Adiabatic band dynamics and high-harmonic pumping}
{Prior to investigating the transport responses of the altermagnetic system, we find it instructive to start with an analysis of the instantaneous band dynamics, which are only well defined within the adiabatic picture where the system’s eigen-states evolve slowly regarding the relevant energy scale, specifically the $s$-$d$ exchange coupling. This approach is motivated by the fact that magnetic dynamics typically operate at frequency scales in the GHz (THz) ranges for ferromagnets (antiferromagnets). This situates magnetic carrier pumping naturally within the adiabatic regime as the $s$-$d$ exchange coupling is typically in the order of the tight-binding hopping parameter. Within this context, one can systematically define adiabatic energy levels and eigen-states \cite{Yue2022}, which are the solutions of the eigenvalue problem for the system’s instantaneous Hamiltonian. To this end, let us  consider a setup in which the magnetic system is contacted to a normal metallic lead (see. Fig. \ref{fig:fig1}). Our primer focus in this section will be on the dynamical part. 
}
Altermagnetic splitting can be described by a tight-binding model \cite{Reichlova2021, Smejkal2022} comprising two contributions: a kinetic nearest-neighbor hopping term $\gamma_0$ and a staggered spin-momentum coupling term parameterized by $\gamma_J$. The corresponding unperturbed Hamiltonian reads 
\begin{equation}
\label{eq:AMmodel}
\mathcal{H}_0 = \sum_{\langle ij \rangle} \hat{c}_i^{\dagger} \left( \gamma_0 + \gamma_J \boldsymbol{\sigma}\cdot \boldsymbol{d}_{ij}\right) \hat{c}_j,
\end{equation}
where  $\boldsymbol{d}_{ij}$ is a unit vector aligned with the N\'eel order, giving rise to an exchange-dependent hopping that is positive (negative) along the $x(y)$ axis.
The quantities $\hat{c}_i$ and $\hat{c}_i^{\dagger}$ denote the annihilation and creation operators at the lattice site $i$, respectively, and $\boldsymbol{\sigma}$ is the vector of Pauli matrices acting in spin space. {The hopping parameter in the normal metal lead will be denoted as $\gamma$ and serves as a unit for the relevant energy scales of the system.}

{In Eq.~\ref{eq:AMmodel}, the term proportional to $\gamma_{J}$ induces the altermagnetic spin-splitting. Since this term represents a hopping contribution, the resulting splitting is inherently momentum-dependent, as shown in the energy dispersions discussed below.
}
Note that this model has been used in a variety of contexts, including optically driven HHG \cite{Werner2024} and the study of giant tunneling and magneto-resistance phenomena in AMs \cite{Smejkal2022}.

\begin{figure*}
    \includegraphics[width=0.8\textwidth]{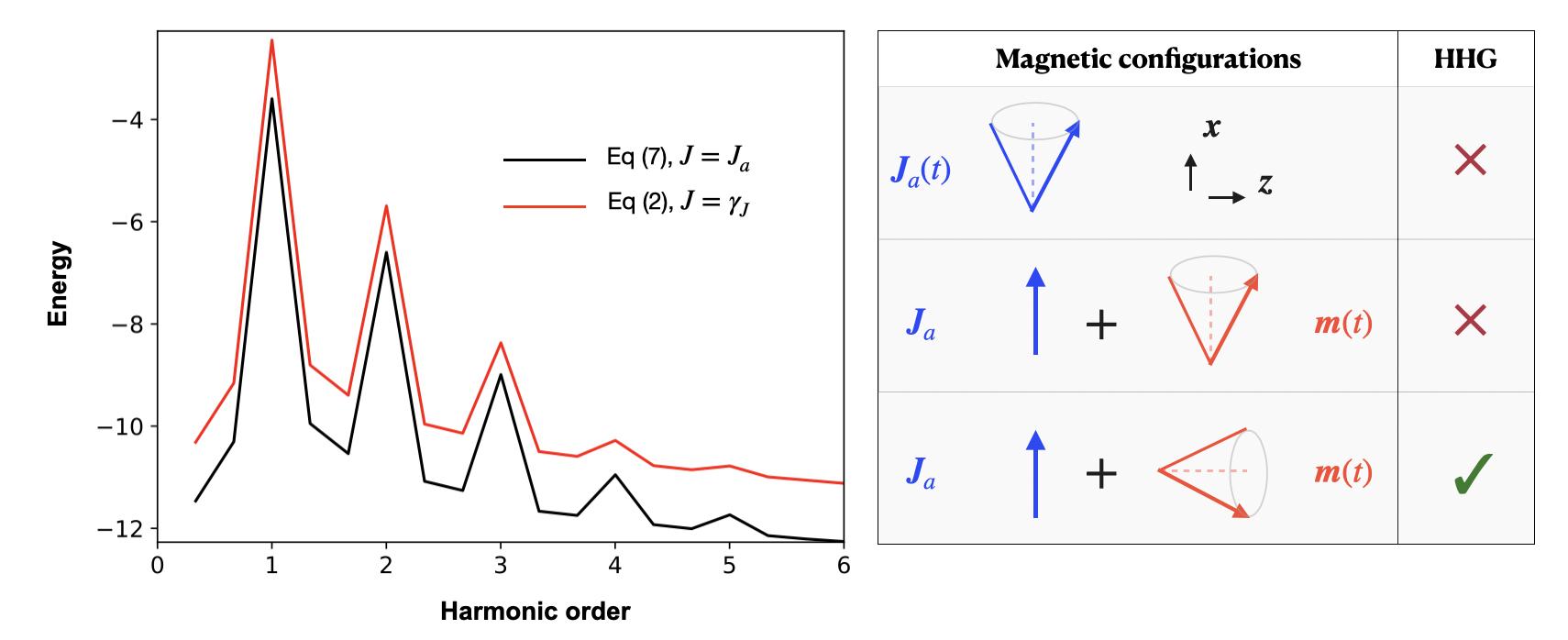}
    \caption{
Left panel: Fourier spectra of the lowest energy level for the simplified and the realistic models [Eqs. \eqref{simple_model} and \eqref{realistic}]. The altermagnetic order parameter is kept static, while the driving magnetic order precesses around the $z$ axis (corresponding to the third configuration shown in the right panel) with a cone angle $\theta=\pi/8$. The instantaneous energy levels are evaluated at $\mathbf{k}=(\pi/2,0)$ for Eq. \eqref{simple_model} and at $\mathbf{k}=(\pi/2,\pi/2,\pi)$—corresponding to the midpoint of the A–Z path in the tetragonal Brillouin zone—for Eq. \eqref{realistic}. The parameters of the realistic model in Eq. \eqref{realistic} correspond to those of RuO$_2$ given in Ref. \cite{Roig2024}. 
Right panel: Selection rules for the relevant magnetic configurations are depicted. HHG occurs only when the precession axis of the driven magnetic order is non-collinear with the ferromagnetic moment.
}
    \label{fig:Fig2}
\end{figure*}

To describe the interaction with a precessing magnetic order, we further add a ferromagnetic exchange interaction of the form $J\mathbf{m}(t)\cdot\boldsymbol{\sigma}$, where
$\mathbf{m}(t)=(\sin\theta\cos\omega t,\;\sin\theta\sin\omega t, \;\cos\theta )$
is a time-dependent unit vector specifying the direction of the magnetization, precessing at angular frequency $\omega$ and cone opening $\theta$. Without loss of generality, we take the N\'eel vector along $x$.
The full time-dependent Hamiltonian then reads
\begin{equation}
\label{simple_model}
\mathcal{H}(t) = \sum_{\langle ij \rangle} \hat{c}_i^{\dagger} \left( {\gamma_0} + \gamma_J \sigma_x \right) \hat{c}_j
+ \sum_i \hat{c}_i^{\dagger} \left( J\,\mathbf{m}(t)\cdot\boldsymbol{\sigma} \right) \hat{c}_i.
\end{equation}

It is important to distinguish two qualitatively different scenarios. First, when the time-dependent magnetization is aligned with the N\'eel order direction, the resulting energy dispersion remains time independent and is given by
\begin{equation}
\label{linearbands}
\varepsilon_\pm(\mathbf{k}) =
f_1(\mathbf{k}) \pm \sqrt{ f_2^2(\mathbf{k}) + J^2 + 2J f_2(\mathbf{k}) \cos\theta },
\end{equation}
where we have introduced the momentum-dependent functions;
\begin{eqnarray}
f_1(\mathbf{k}) &= 2\gamma_0 \left( \cos k_x + \cos k_y \right), \\
f_2(\mathbf{k}) &= 2\gamma_J \left( \cos k_x - \cos k_y \right).
\end{eqnarray}
In this configuration, the dynamics does not lead to a nonlinear response.

Second, when the magnetization precesses around an axis perpendicular to the N\'eel order, 
the bulk energy dispersion acquires an explicit time dependence and takes the form
\begin{equation}
\label{bands}
\varepsilon_\pm(\mathbf{k}, t) =
f_1(\mathbf{k}) \pm \sqrt{ f_2^2(\mathbf{k}) + J^2
+ 2J f_2(\mathbf{k}) \sin\theta \cos(\omega t) }.
\end{equation}

According to our recent proposal \cite{Ly2025}, only energy dispersions that exhibit a nonlinear time dependence can give rise to HHG. We therefore focus on this second scenario and fix the precession axis along a direction perpendicular to the N\'eel order. 

Although the nonlinear nature of the energy levels is transparent in the simple two-dimensional model discussed above [Eq. \eqref{bands}], analytical expressions for the spectrum are generally unavailable in more realistic settings. To gain further insight into the selection rules underlying HHG in AMs, we therefore numerically investigate the band dynamics of a generic three-dimensional model introduced in Ref. \cite{Roig2024}. This model reproduces density-functional-theory calculations for a variety of altermagnetic systems, including the tetragonal AM RuO$_2$. Without loss of generality, we focus on the single-orbital version of the model; calculations based on the two-orbital model (proposed in the same reference) yield similar nonlinear band dynamics.

\begin{figure*}
    \includegraphics[width=\textwidth]{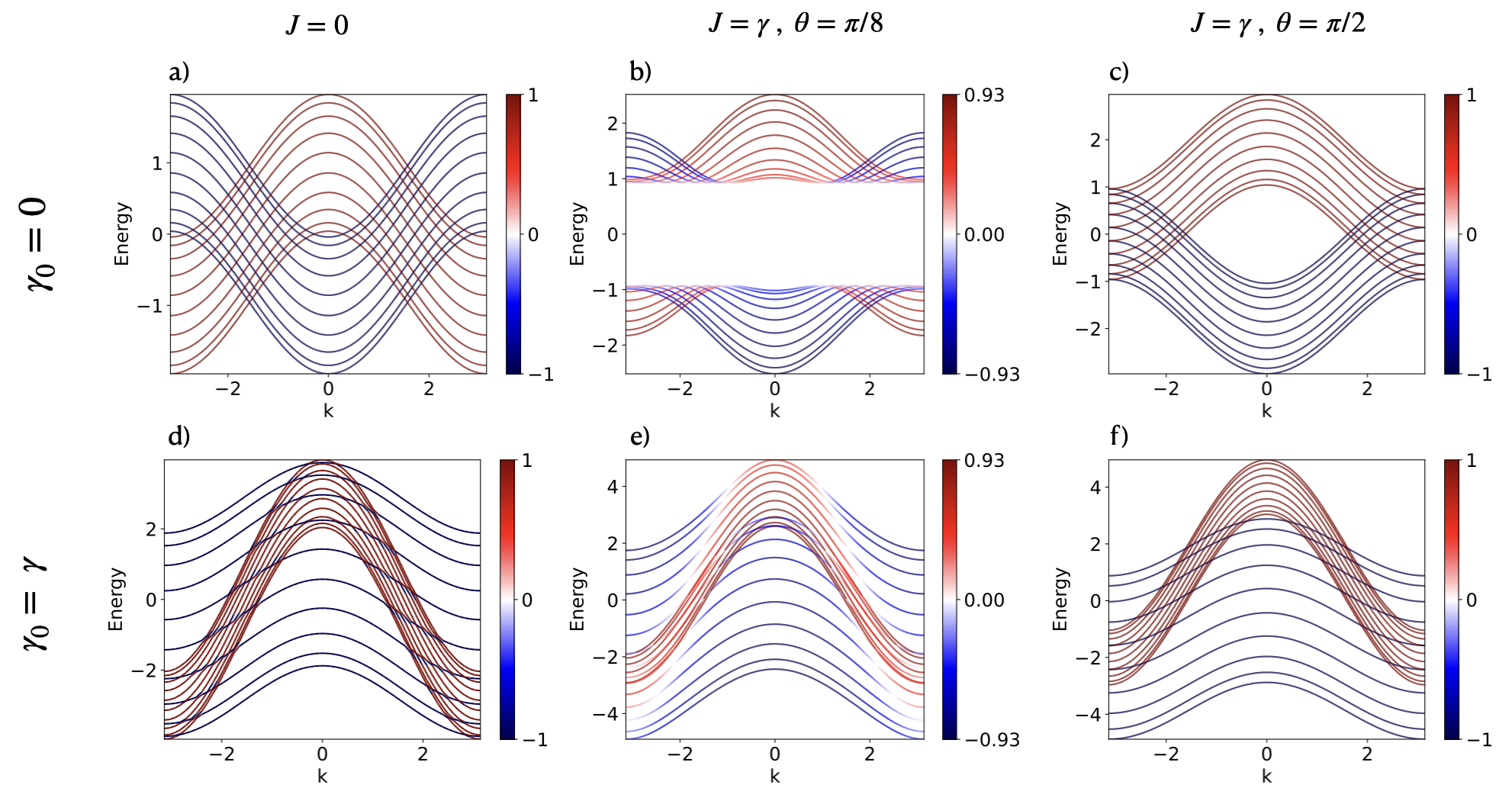}
    \caption{
Altermagnetic band structures for different magnetic configurations. The first column shows the band structure in the absence of magnetic dynamics. The second column displays the spin-polarized bands for a precession cone angle, $\theta=\pi/8$. The third column shows the bands for fully in-plane dynamics, $\theta=\pi/2$. In the upper row, the hopping parameter is set to $\gamma_0=0$, while in the lower row it is set to $\gamma_0=\gamma$. In all panels, the Hamiltonian is diagonalized at time $t=0$.
}
    \label{fig:Fig3}
\end{figure*}

The underlying single-orbital Hamiltonian reads
\begin{equation}
\label{realistic}
\mathcal{H}(t) = \varepsilon_0(\mathbf{k}) + t_{x\mathbf{k}} \tau_x + t_{z\mathbf{k}} \tau_z
+ \tau_z \boldsymbol{J}_a \cdot \boldsymbol{\sigma} + J\,\mathbf{m}(t)\cdot \boldsymbol{\sigma},
\end{equation}
where $\varepsilon_0(\mathbf{k})$, $t_{x\mathbf{k}}$, and $t_{z\mathbf{k}}$ are momentum-dependent terms specified in Ref. \cite{Roig2024}. The vector $\boldsymbol{J}_a$ denotes the altermagnetic primary order parameter, which we take to be oriented along the $x$ direction. The time-dependent magnetization $\mathbf{m}(t)$, with coupling strength $J$, represents the driving magnetic dynamics added to the model. The $2\times2$ matrices $\boldsymbol{\tau}$ and $\boldsymbol{\sigma}$ denote the Pauli matrices acting in sub-lattice and spin spaces, respectively. 
{Note that a tensor product between these two matrices is implicitly assumed. The scalar term is treated as being multiplied by the $4 \times 4$ identity matrix. 
Terms proportional to $\tau$ ($\sigma$) denote tensor products with the 
$2 \times 2$ identity matrix acting on the right (left) subspace respectively.}
The relativistic SOC contribution is totally omitted. In the following, we analyze the dynamics of Eq. \eqref{realistic} numerically, without resorting to analytical diagonalization.

Figure \ref{fig:Fig2} shows the Fourier spectrum of the lowest energy level obtained from numerical diagonalization of the three-dimensional model Hamiltonian [Eq. \eqref{realistic}], alongside the corresponding spectrum of the two-dimensional model [Eq. \eqref{bands}]. In both cases, similar higher harmonics emerge in the energy levels, highlighting the robustness of the underlying mechanism. 
{Importantly, however, the appearance of high harmonics obeys	specific selection rules.
The advantage of the adiabatic band dynamics analysis, is that it enables us to anticipate these selection rules in different emission scenarios without resorting to tedious non-equilibrium transport calculations.}

Two distinct driving protocols can be considered: either the altermagnetic order parameter itself is time-dependent, or an adjacent magnetic order (ferromagnetic or antiferromagnetic) is driven. For simplicity, we focus on the ferromagnetic case. If the altermagnetic order precesses around the $z$ axis in the absence of any secondary magnetic order, no HHG occurs, and the resulting energy dispersion remains linear, similarly to Eq. \eqref{linearbands}. When the altermagnetic order is static and the adjacent ferromagnetic order precesses around the same axis, the spectrum again remains linear. High harmonics emerge only when the driving magnetic order precesses around an axis that is non-collinear with the altermagnetic order parameter. These scenarios are fully confirmed by our numerical analysis and are summarized in the right panel of Fig. \ref{fig:Fig2}. This establishes the generality of the mechanism and clarifies the conditions required for HHG in purely magnetic systems.

In the following section, we turn to the transport properties and investigate {altermagnetic spin and charge pumping} using the simplified model introduced in Eq. \eqref{simple_model}, {with a guiding provided by the adiabatic band dynamics analysis given above.}

\begin{figure*}
    \includegraphics[width=0.8\textwidth]{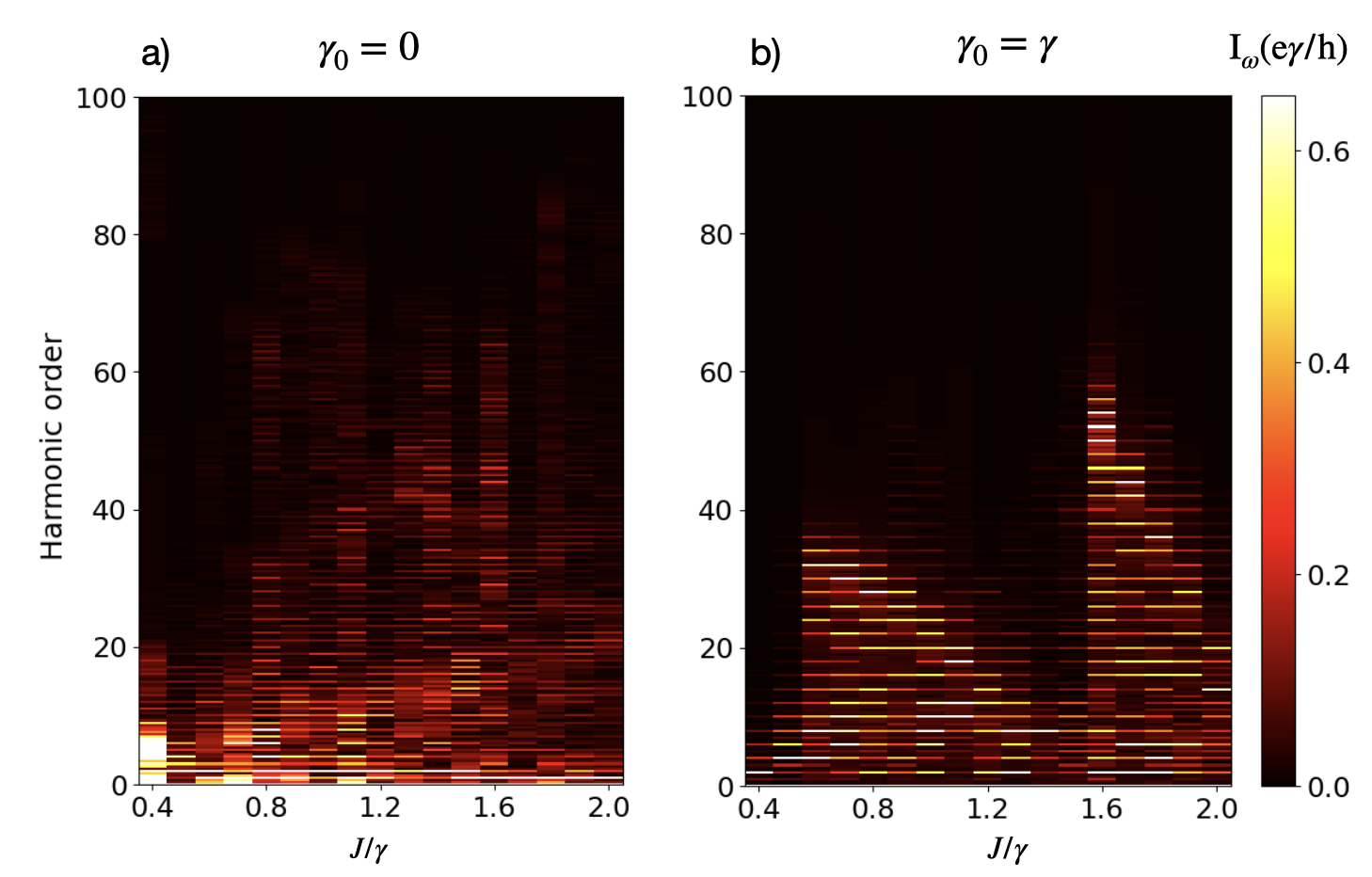}
    \caption{
Fourier spectrum of the charge current pumped from the AM. The driving frequency is set to $\omega=0.01\,\gamma/h$. The Fermi energy is chosen as $\varepsilon_{\rm F}=0$ for $\gamma_0=\gamma$ and $\varepsilon_{\rm F}=-J$ for the gaped case ($\gamma_0=0$). The precession cone angle is $\theta=\pi/8$. The data are represented on a linear scale.
}
    \label{fig:Fig4}
\end{figure*}

\section{Numerical Results}

To quantify spin and charge pumping in altermagnetic systems, we carry out numerical simulations of the non-equilibrium transport response arising from the coupling between a driven ferromagnetic order and an altermagnetic background. The calculations are based on the tight-binding Hamiltonian introduced in the previous section. In the context of magnetically induced HHG, three parameters govern the dynamics: the excitation frequency $\omega$, the exchange amplitude $J$, and the spin-splitting strength, parameterized by the altermagnetic coupling $\gamma_J$, in addition to the cone opening $\theta$.

In the simulations presented below, we fix $\gamma_J = 0.5\gamma$ and choose a driving frequency $\omega = 0.01\,\gamma/h$. The normal hopping parameter in the AM is taken as $\gamma_0=0$ or $\gamma_0=\gamma$, corresponding to different spin-splitting scenarios. The resulting altermagnetic band structures for both cases are shown in Fig. \ref{fig:Fig3}. When the magnetic dynamics is switched on (second column), the altermagnetic bands become gaped due to the ferromagnetic exchange interaction. For $\gamma_0 = 0$, a clear band splitting of magnitude $2J$ is observed for a small precession angle $\theta = \pi/8$. By contrast, for $\gamma_0 = \gamma$, the interplay between the altermagnetic order and the ferromagnetic moment results in a more subtle band splitting, as shown in Fig. \ref{fig:Fig3}(c). Depending on the relative orientations of the two magnetic orders, distinct splitting scenarios can emerge.

{
We now turn to the calculation of the nonlinear transport response. To this end, we consider a rectangular scattering region of width $W = 10a$ and length $L = 20a$, connected to a normal lead (see Fig. \ref{fig:fig1}). We focus on the configuration in which the altermagnetic order parameter is oriented along the $x$ direction, while an adjacent ferromagnetic order precesses around the $z$ axis. This corresponds to the non-collinear magnetic alignment illustrated in Fig.~\ref{fig:Fig2}. The magnetic moments are allowed to precess in time, and the resulting charge current flowing into the normal lead is computed.

In AMs, non-relativistic spin--momentum couplings allow time-dependent magnetic order to induce charge currents through spin--charge conversion mechanisms \cite{Kokkele2025}. Consequently, magnetic precession generally produces a finite AC charge response, whose temporal structure reflects the nonlinear nature of the underlying spin dynamics. In the one-terminal geometry considered here, however, charge conservation enforces the absence of any DC component in the periodic steady state. Nonetheless, the AC charge current provides a direct probe of the nonlinear and high-harmonic content of the transport response. 
While the present setup does not support DC charge pumping, the spin currents generated by the precessing dynamics may be converted into measurable transverse voltages in multi-terminal detection schemes based on inverse spin Hall effects \cite{Dou2025, Dongchao2026, Costache2006, Jeon2018} or inverse Edelstein effects \cite{Chakraborty2025, Pal2023}.

Figure~\ref{fig:Fig4} shows the Fourier spectrum of the pumped charge current for a precession cone angle $\theta=\pi/8$, for both $\gamma_0=0$ and $\gamma_0=\gamma$. In both cases, we observe the emission of approximately one hundred harmonics with amplitudes comparable to that of the fundamental frequency. Higher-order harmonics beyond the 100th order are also present, albeit with rapidly decreasing amplitudes. While these components form a tail in the spectrum, we focus here on the most pronounced harmonics, as they constitute the key qualitative distinction between magnetically driven high-harmonic generation and its light-driven counterpart.
}

\begin{figure*}
    \includegraphics[width=0.8\textwidth]{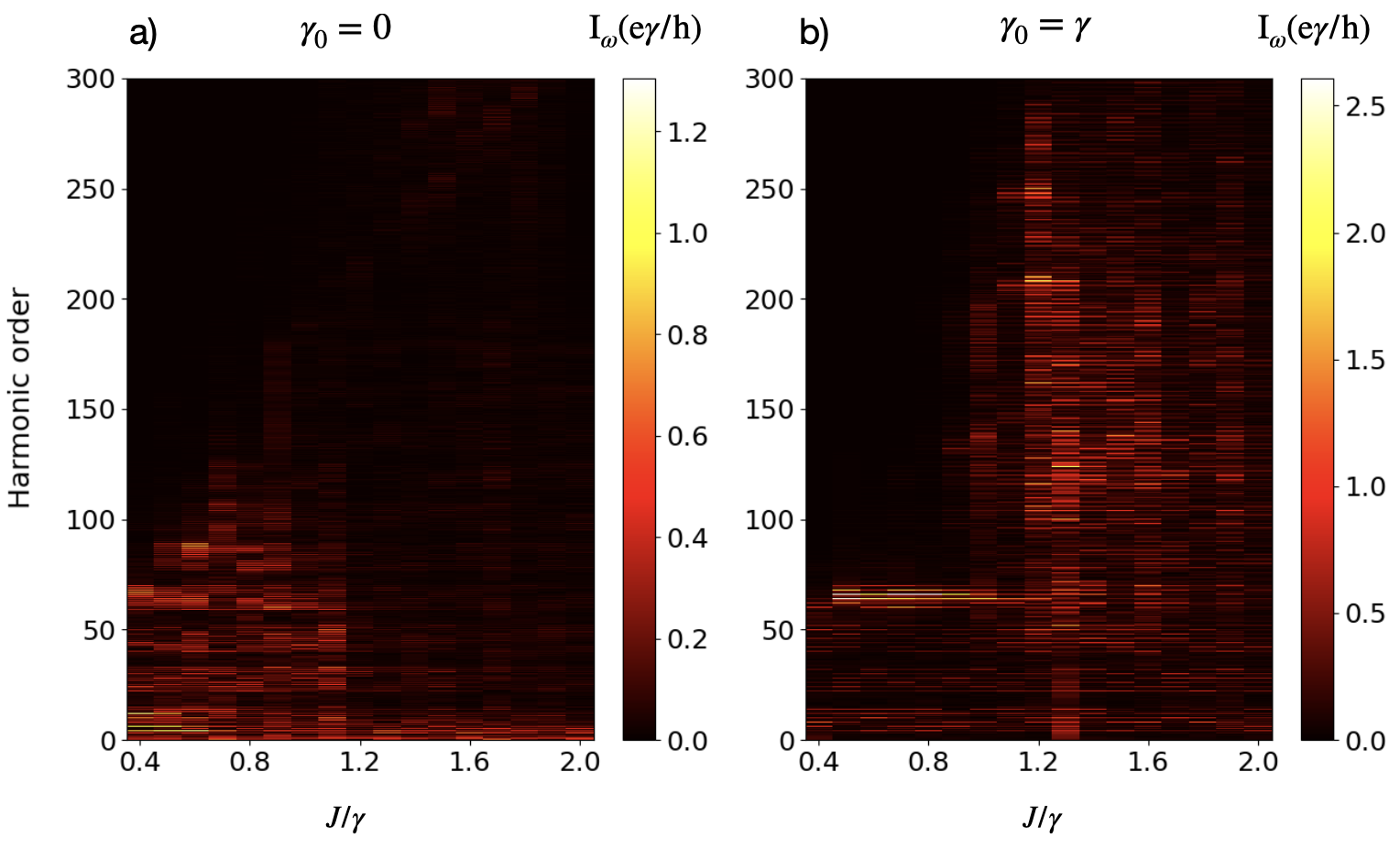}
    \caption{
Fourier spectrum of the charge current pumped out of the AM for fully in-plane magnetic dynamics, corresponding to a precession cone angle $\theta=\pi/2$. All other parameters are identical to those used in Fig. \ref{fig:Fig4}. 
The data are represented on a linear scale.
}
    \label{fig:Fig5}
\end{figure*}

Although ferromagnetic dynamics is typically associated with small precession cone angles, large-angle dynamics is of considerable theoretical interest. In particular, fully in-plane dynamics corresponds to a maximal coupling between the electronic spins and the precessing magnetic order. The resulting charge emission for this configuration is shown in Fig. \ref{fig:Fig5}. In this regime, we observe a strongly enhanced harmonic response extending up to the 300th order. These high-order harmonics are more pronounced for $\gamma_0=\gamma$, but remain clearly observable in the case $\gamma_0=0$, albeit with reduced amplitudes.

It is worth emphasizing that magnetically driven HHG often exhibits strong enhancement near specific resonance conditions, as previously reported for systems with relativistic SOC \cite{Ly2022} {with particular scaling laws \cite{Ly2025b}, different from the optical driving case \cite{Krause1992}.} In contrast, the present results reveal a broad region of parameter space in which the nonlinear response is strongly enhanced. In the altermagnetic system considered here, the momentum dependence of the instantaneous band structure [Eq. \eqref{bands}] is significantly more complex than in Rashba-type systems, making it difficult to identify a simple resonance condition governing the enhancement of higher harmonics. By comparison, in the Rashba case the resonance condition reduces to the simple relation $J = k\alpha$, where $\alpha$ denotes the SOC strength and $k$ the Fermi wave vector.

{While the band dynamics perspective identifies an elementary mechanism for magnetically driven HHG, a more phenomenological picture can be adopted. Specifically, non-relativistic altermagnetic spin-splitting leads to the emergence of spin-flip scattering events within each magnetic cycle. Drawing an analogy to the well-known three-step model \cite{Corkum1993, Lewenstein1994,Ghimire2026} of optically driven HHG, the harmonic band-width would simply correspond to the effective number of spin-flip scattering events occurring within a single driving period.}

An essential point to highlight concerns the excitation of the full electronic bandwidth in the in-plane (or nearly in-plane) dynamical configuration. Although this regime remains primarily of theoretical interest and poses experimental challenges, several viable pathways toward its realization have been explored. In particular, large-angle precession can be stabilized using oscillating in-plane magnetic fields \cite{Watts2006,Jia2024} or by applying appropriately optimized microwave frequencies \cite{Yamamoto2020,Imamura2020}. Complementary approaches based on spin-transfer-torque and spin–Hall-torque nano-oscillators also provide direct access to large-angle dynamical regimes \cite{Kiselev2003,Krivorotov2008,Tingsu2016,Sheng2024}.

This perspective is especially appealing, as it provides a realistic platform for enhancing strong nonlinear responses and the associated spin and charge pumping in a broad class of spintronic devices, including the setup considered in the present work.

\begin{figure}
    \centering
    \includegraphics[width=\linewidth]{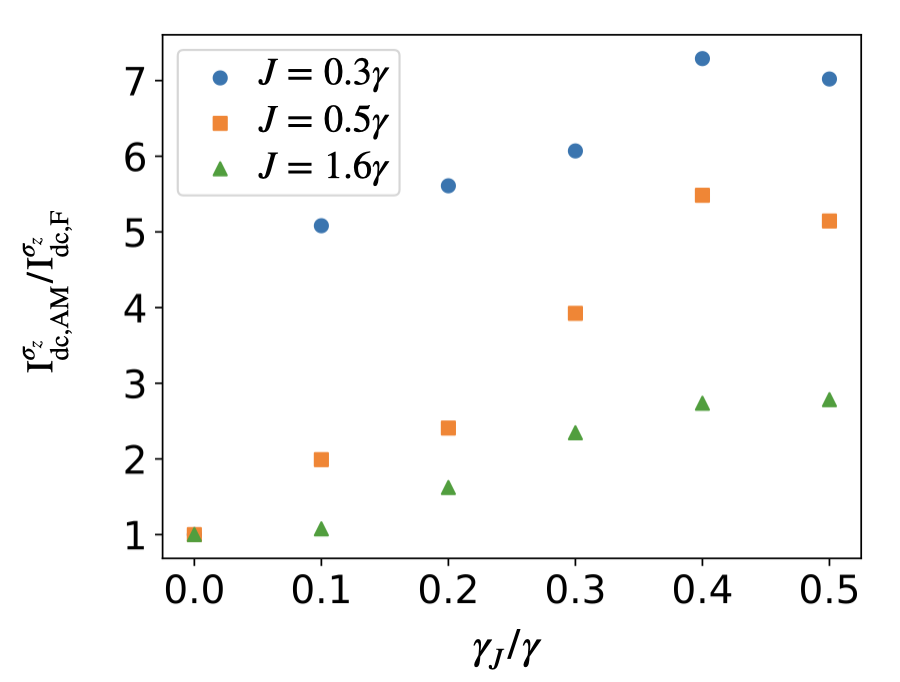}
    \caption{The dependence of the DC pumped spin current on the altermagnetic parameter $\gamma_J$ for different $s$-$d$ coupling strengths is shown. The DC amplitude is normalized to the reference ferromagnetic amplitude (obtained at $\gamma_J=0$). Parameters identical to those of Fig. \ref{fig:Fig4}(b) were used.}
    \label{fig:Fig6}
\end{figure}

{Finally, it is important to quantify the magnitude of the expected DC spin current from the altermagnetic system. Without loss of generality, we consider the  spin current polarized along the $z$ direction.  
In Fig. \ref{fig:Fig6}, the DC component of the spin current is shown as a function of $\gamma_J$, for different $s$-$d$ exchange coupling values. For a quantitative comparison with the ferromagnetic pumping, the current is normalized by the ferromagnetic DC pumping amplitude, that is in the absence of the altermagnetic order ($\gamma_J=0$). One clearly observes an enhancement of the DC current as the altermagnetic parameter is increased. A strong enhancement of pumping is observed at low $J$. 
{For larger coupling strengths, the magnetic gap increases; this hinders the transmission of propagating transport modes and reduces the DC spin current amplitude. A similar suppression is observed at large $\gamma_J$. This demonstrates that the reported enhancement of pumping is sensitive to parameter tuning. Comparable enhancement trends have been observed in AMs \cite{Sun2023}, upon appropriately tuning the relativistic SOC strength; in contrast, in our setup the enhancement is rather induced by the non-relativistic splitting. Ultimately, these findings suggest that the effect can be efficiently detected using the experimental techniques established for probing ferromagnetic pumping.}
}

\section{Discussion and Conclusion}
Our results demonstrate that altermagnetic systems can host strongly nonlinear {spin and charge currents} when coupled to magnetic dynamics. While the intrinsic altermagnetic resonance alone corresponds to a gauge-trivial regime with effectively time-independent band dispersions, the introduction of an additional magnetic order restores a genuine nonlinear time dependence. In this regime, magnetically driven dynamics gives rise to high-harmonic emission extending to several hundred harmonics with sizable amplitudes.

A comparison with systems exhibiting relativistic spin splitting highlights a key advantage of the present non-relativistic platform. In Rashba-type systems, access to the ultra-high-harmonic regime typically requires exceptionally large SOC strengths. By contrast, in altermagnetic systems the strongly nonlinear regime can be reached with moderate $s$--$d$ exchange and alternating spin-momentum coupling. This renders AMs particularly attractive for applications requiring efficient generation of high-frequency signals, including terahertz emitters and nonlinear spintronic devices.

An additional advantage lies in the realization of altermagnetic states in bulk three-dimensional materials, which significantly facilitates experimental implementation. This contrasts with mechanisms relying on relativistic SOC, where the relevant interactions are often confined to interfaces or surfaces, thereby imposing additional constraints on material design and device fabrication.
{Different magnetic materials have been suggested to host the altermagnetic spin-splitting, including RuO$_2$ \cite{Berlijn2017,Ahn2019,Smejkal2020}, MnF$_2$ \cite{Yuan2020, Bhowal2024}, MnTe \cite{Lee2024} and other materials \cite{Naka2019,Mazin2021,Reimers2024, Fedchenko2024,Reimers2024,Jianyang2024, Yang2025,Yang2025b}.}

From the perspective of {carrier pumping}, the driving magnetic dynamics may be excited via altermagnetic resonance \cite{Gomonay2024}, closely analogous to antiferromagnetic resonance \cite{Cheng2014,Vaidya2020}. In this scenario, the alternating spin texture itself undergoes dynamical motion, with a characteristic frequency naturally lying in the terahertz range, reflecting the intrinsic dynamics of the N\'eel order in compensated collinear magnets. However, this corresponds to a regime in which the time dependence can be removed by an appropriate gauge transformation, resulting in effectively time-independent band dispersions [Eq. \eqref{linearbands}]. Genuine nonlinear time dependence, and hence {strongly nonlinear pumping}, can be restored by introducing an additional  transverse magnetic field perpendicular to the altermagnetic order. Importantly, this transverse component need not be dynamical \cite{Ly2026}; a static field is sufficient to prevent the removal of the time dependence, leading to an energy dispersion of the form given in Eq. \eqref{bands}. Alternatively, nonlinear dynamics may also arise through coupling to another driven magnetic order, which can be of either ferromagnetic or antiferromagnetic nature, corresponding to the third configuration illustrated in Fig. \ref{fig:Fig2}.

{While altermagnetic systems have already been proposed as a platform for light-driven HHG \cite{Werner2024}, the present results indicate that combining optical excitation with magnetic dynamics provides an appealing route toward the experimental observation of magnetically induced high-harmonic {spin and charge} pumping using purely optical techniques. Moreover, the broad parameter window over which strong nonlinear emission is observed relaxes the need for fine-tuned resonance conditions, thereby enhancing experimental robustness.}

{
Within this context, it is instructive to comment on the role of the driving frequency and its implications for the generality of our results. In numerical terms, taking a hopping parameter $\gamma = 0.5 \mathrm{eV}$ corresponds to a driving frequency $\omega \simeq 1.25 \mathrm{THz}$, which is characteristic of antiferromagnetic (or altermagnetic) resonance. Under these conditions, the present mechanism can enhance the dynamical response by up to two orders of magnitude, enabling ultrafast dynamics extending beyond $100 \mathrm{THz}$. 

Although such frequencies are employed here in the context of a ferromagnetic precession, the underlying mechanism is not restricted to this choice. Indeed, while ferromagnetic resonance typically occurs in the GHz range, our results remain qualitatively unchanged. To substantiate this point, we present in the Supplemental Materials additional simulations in which a compensated antiferromagnetic dynamics is considered instead. We find that the HHG features persist in this case, confirming the generality of our conclusions. The use of a ferromagnetic precession at THz frequencies therefore does not alter the generality of our results.

Rather, this analysis naturally highlights the importance of the frequency dependence of the effect. In conventional spin pumping, the DC spin current scales linearly with the driving frequency $\omega$. By contrast, in the present mechanism the DC component of the pumped response exhibits a strongly non-monotonic dependence on $\omega$, reflecting the highly nonlinear character of the underlying dynamics (see Supplemental Materials).
}

{The non-monotonic frequency dependence discussed above is  therefore a clear signature of the underlying nonlinear dynamics.}

From a device perspective, the non-relativistic origin of the effect is particularly appealing. Unlike schemes relying on strong relativistic SOC—which are often limited to interfaces or surface states—the present mechanism is compatible with bulk three-dimensional altermagnetic materials. This opens the door to scalable architectures for compact terahertz emitters, frequency multipliers, and nonlinear spintronic components. The ability to control the harmonic spectrum through magnetic proximity or excitation geometry provides additional flexibility for device optimization.

{We anticipate that the proposed effect can be experimentally probed using inverse spin Hall setups, where the pumped spin current is converted into a transverse voltage \cite{Saitoh2006,Mosendz2010}. While measuring the time-dependent signal may prove experimentally challenging, the nonlinear band dynamics mirroring the HHG effect suggests that it could be signaled via angle resolved time-domain photoemission spectroscopy \cite{Zhong2022}. Furthermore, the interplay between magnetic and optical driving offers a promising route to observe the effect through purely optical means.}

In conclusion, building upon realistic modeling, we have demonstrated that altermagnetic systems can exhibit genuinely nonlinear dynamical responses in both spin and charge pumping in the absence of relativistic SOC. By analyzing the instantaneous energy levels in the adiabatic regime, we identified the highly nonlinear character of spin and charge pumping driven by magnetic dynamics in AMs. Furthermore, our systematic investigation of the magnetic configurations establishes the fundamental selection rules determining {when spin and charge pumping in AMs} acquires a strongly nonlinear character. These results identify AMs as a promising platform for ultrafast spintronic functionalities, opening new routes toward high-frequency {carrier} control based solely on magnetic structure and dynamics.

\acknowledgments
We thank A. Abbout, P. Fu and Satoru Hayami for useful discussions. 

\bibliography{refs}
\end{document}